\documentclass[10pt,conference]{IEEEtran}
\usepackage[left=0.63in,right=0.63in,top=0.7in,bottom=1.05in]{geometry}

\usepackage[dvipsnames]{xcolor}
\usepackage[normalem]{ulem}
\DeclareRobustCommand{\erase}{\bgroup\markoverwith{\textcolor{red}{\rule[.5ex]{2pt}{0.5pt}}}\ULon}

\newcommand*{\tran}{\mathsf{T}}
\newcommand*{\herm}{\mathsf{H}}

\usepackage{cite}

\usepackage{graphicx}
\graphicspath{{./figures/}}
\DeclareGraphicsExtensions{.pdf,.jpeg,.png}

\usepackage{amsmath,amssymb,amsthm,amsfonts,amscd,amsbsy,amsxtra,amsfonts}
\usepackage{mathtools}

\usepackage[ruled]{algorithm2e}
\SetKwFor{For}{for}{}{end}
\SetKwFor{For}{for}{}{end}
\SetKwFor{If}{if}{}{end}

\usepackage{booktabs}
\usepackage{multirow}

\usepackage[caption=false,font=footnotesize]{subfig}

\DeclareMathOperator{\diag}{diag}
\DeclareMathOperator*{\argmax}{arg\,max}

\DeclareMathOperator{\var}{Var}

\DeclarePairedDelimiter\abs{\lvert}{\rvert}%
\DeclarePairedDelimiter\norm{\lVert}{\rVert}%

\newtheorem{definition}{\underline{Definition}}

\IEEEoverridecommandlockouts

\begin{document}
%
\title{Orthogonal Delay-Doppler Division Multiplexing Modulation with Hierarchical Mode-Based Index Modulation}

\author{
    \IEEEauthorblockN{
        Kehan Huang,
        Min Qiu,
        Jinhong Yuan,
    }
    \IEEEauthorblockA{
        University of New South Wales, Australia, Email: \{kehan.huang, min.qiu, j.yuan\}@unsw.edu.au
    }
    \thanks{This work was supported in part by the Australian Research Council (ARC) Discovery Project under Grant DP220103596, and in part by the ARC Linkage Project under Grant LP200301482.}
}

\maketitle

\begin{abstract}
    The orthogonal time frequency space with index modulation (OTFS-IM) offers flexible tradeoffs between spectral efficiency (SE) and bit error rate (BER) in doubly selective fading channels. While OTFS-IM schemes demonstrated such potential, a persistent challenge lies in the detection complexity. To address this problem, we propose the hierarchical mode-based index modulation (HMIM). HMIM introduces a novel approach to modulate information bits by IM patterns, significantly simplifying the complexity of maximum a posteriori (MAP) estimation with Gaussian noise. Further, we incorporate HMIM with the recently proposed orthogonal delay-Doppler division multiplexing (ODDM) modulation, namely ODDM-HMIM, to exploit the full diversity of the delay-Doppler (DD) channel. The BER performance of ODDM-HMIM is analyzed considering a maximum likelihood (ML) detector. Our numerical results reveal that, with the same SE, HMIM can outperform conventional IM in terms of both BER and computational complexity. In addition, we propose a successive interference cancellation-based minimum mean square error (SIC-MMSE) detector for ODDM-HMIM, which enables low-complexity detection with large frame sizes.
\end{abstract}


\section{Introduction}\label{sec:introduction}
With the advancement of technologies like high-speed rail and communication satellites, the demand for high-mobility communication is growing increasingly important in the next-generation wireless networks. One of the challenges in such environments is the severe time selectivity of the channel caused by the Doppler effect. In light of this, the orthogonal time frequency space (OTFS) modulation is proposed to modulate information bits in the delay-Doppler (DD) domain, where the channel has nice properties \cite{Hadani2017OrthogonalModulation}. However, due to the uncertainty principle, the ideal bi-orthogonal pulse for OTFS does not exist \cite{Raviteja2018InterferenceModulation,Lin2022OrthogonalModulation}. Therefore, rectangular pulses are usually considered in the literature \cite{Raviteja2018InterferenceModulation,Thaj2020LowSystems}, which leads to high out-of-band emission (OOBE) \cite{Lin2022OrthogonalModulation} and complicated ISI \cite{Lin2022OrthogonalModulation,Tong2024ODDM_PhyChan}. To address these issues, the orthogonal delay-Doppler division multiplexing (ODDM) modulation has been recently proposed. Featuring the practically realizable delay-Doppler plane orthogonal pulse (DDOP), ODDM ensures sufficient orthogonality on the DD plane while maintaining low OOBE \cite{Lin2022OrthogonalModulation,Lin2022OnPulse}. Without the complicated ISI from rectangular pulses, ODDM enjoys an \emph{exact} DD domain input-output relation. This allows ODDM receivers to perform more accurate signal detection. The performance of ODDM has been studied in \cite{Huang2024ODDM_Performance}. In this paper, we adopt ODDM modulation and its input-output relation to enable DD domain communications.

Index modulation (IM) was first introduced by orthogonal frequency-division multiplexing with IM (OFDM-IM), where part of the information bits are conveyed by the patterns of subcarrier indexes \cite{Basar2013OrthogonalModulation}. The indexes are grouped into blocks and the pattern of each block carries information bits according to a pre-defined one-to-one mapping. The fusion of IM with DD domain communication technologies, such as OTFS with IM (OTFS-IM) \cite{Liang2020DopplerModulation} and its variants \cite{Zhao2021OTFS_DMIM,Qian2023OTFS_BlockwiseIM}, leverages this principle in the DD domain. By exploiting the indexes of transmission entities, IM provides an extra dimension to transfer information, which enhances spectral efficiency (SE) and enables flexible tradeoffs with bit error rate (BER).

Despite the benefit in SE, OTFS-IM places additional challenges to receiver design. When linear equalizers are considered, classic OTFS or OFDM systems without IM perform maximum likelihood (ML) detection efficiently in a symbol-wise manner. For systems with IM, however, block-wise ML detection is necessary, where the computational complexity increases exponentially with block length. Therefore, the symbol-wise log-likelihood ratio (LLR) detector \cite{Basar2013OrthogonalModulation,Zhang2017OFDM_DMIM}, modified LLR detector \cite{Zhao2021OTFS_DMIM}, and energy-detection (ED) detector \cite{Zhang2017OFDM_DMIM} are proposed to reduce detection complexity at the cost of increased BER. When considering nonlinear detectors, such as expectation propagation (EP) \cite{Li2022OTFS_IM_EP} and the message passing algorithm (MPA) \cite{Qian2023OTFS_BlockwiseIM}, the challenge in complexity persists due to the block-wise computation of probabilities.

In this paper, we propose a novel IM scheme, namely the ODDM with hierarchical mode-based index modulation (ODDM-HMIM), to address the receiver complexity problem. ODDM-HMIM features a hierarchical QAM constellation (HQC), which effectively integrates multiple modes of a regular QAM constellation. The patterns formed by modes facilitate a simplified IM method to convey extra information bits. A BER analysis is performed for ODDM-HMIM in fading channels. We then derive the maximum a posteriori (MAP) estimator for ODDM-HMIM, which realizes linear computational complexity with respect to the block length. Our numerical results indicate that, compared to ODDM and ODDM with IM (ODDM-IM), ODDM-HMIM offers comparable or better BER performance, while exhibiting low complexity. With the MAP estimator, we further introduce a modified successive interference cancellation-based minimum mean square error (SIC-MMSE) detector. By exploiting the structure of the time-domain ODDM input-output relation, SIC-MMSE markedly reduces the detection complexity of ODDM-HMIM in time-varying channels, even with large frame sizes. In addition, by iteratively harvesting MAP estimation gains, SIC-MMSE achieves a substantial improvement in BER over conventional linear MMSE detectors. Notably, the proposed HMIM scheme is also applicable to OTFS modulation. As the performance of ODDM and OTFS has been compared in \cite{Lin2022OrthogonalModulation}, this paper focuses on ODDM as a practical realization of DD modulation.

\section{System Model}

In this section, we present the ODDM-HMIM scheme. Consider a delay-Doppler multi-carrier (DDMC) signal spanning a duration of $NT$, where $N$ subcarriers are spaced by $\frac{1}{NT}$ and $M$ multi-carrier symbols are staggered by $\frac{T}{M}$. The signal can be discretized on an $M\times N$ DD gird with a time resolution of $\mathcal{T}_\mathrm{DD}=\frac{T}{M}$ and a frequency resolution of $\mathcal{F}_\mathrm{DD}=\frac{1}{NT}$. $MN$ information-bearing symbols are arranged in the grid and form a transmitted frame $\mathbf{X}\in\mathbb{C}^{M\times N}$. 

\vspace{-2mm}
\begin{figure}[ht]
    \centering
    \includegraphics[width=8.6cm,trim={18 0 14 28},clip]{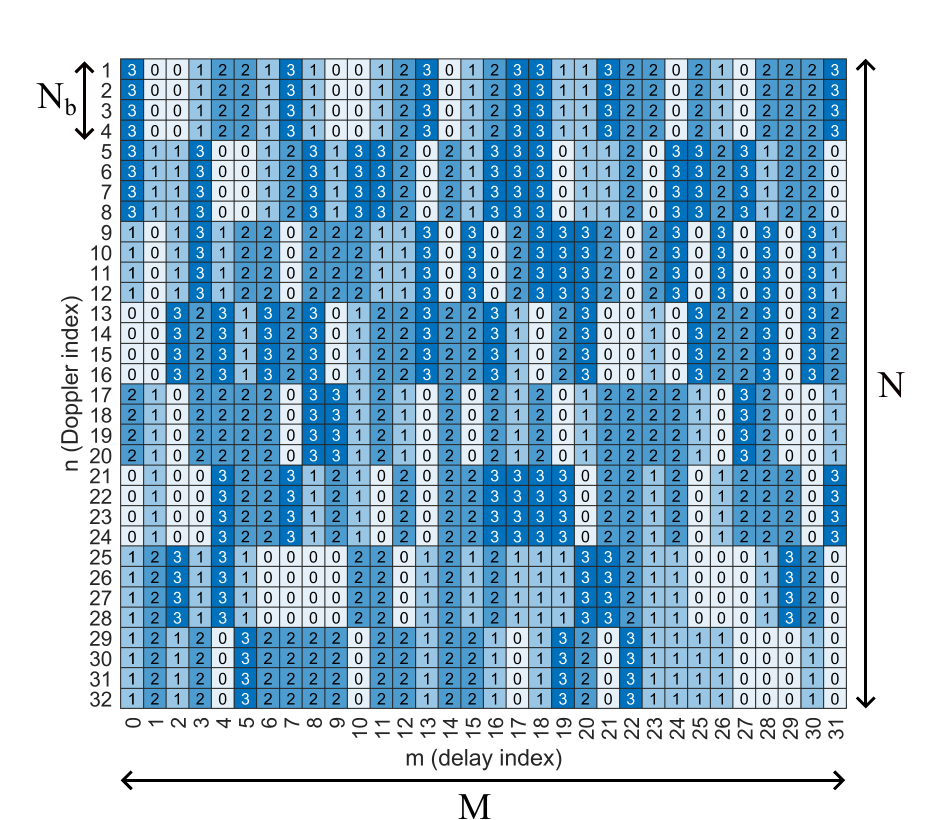}
    \vspace{-3mm}
    \caption{ODDM-HMIM frame structure with $M=N=32,N_b=4$, and $Q_2=4$.}
    \label{fig:frame}
    \vspace{-3mm}
\end{figure}

In the ODDM-HMIM framework, $\mathbf{X}$ is organized into blocks in the Doppler domain with a length of $N_b<N$, as illustrated in Fig. \ref{fig:frame}. To enable the SIC-MMSE detector in section \ref{sec:sicmmse}, each block extends along the Doppler dimension. Let $\mathbf{x}_m\in\mathbb{C}^{N\times1}$ denote the transmitted symbol vector at the $m$-th delay index, for $m\in\{0,\dots,M-1\}$. Then, we define the $\beta$-th block in $\mathbf{x}_m$ as $\mathbf{x}_{m,\beta} = \left[x_m[\beta N_b],\dots,x_m[(\beta+1) N_b -1]\right]^\tran$, for $\beta\in\{0,\dots,N/N_b-1\}$ with $N/N_b\in\mathbb{Z}$.

\subsection{ODDM-HMIM Transmitter}
\begin{figure}[ht]
    \centering
    \includegraphics[width=8.5cm,trim={0 0 0 0},clip]{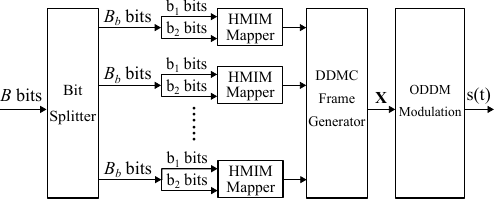}
    \vspace{-2mm}
    \caption{Block diagram of the transmitter of ODDM-HMIM system.}
    \label{fig:block_diagram}
    \vspace{-2mm}
\end{figure}

The structure of the ODDM-HMIM transmitter is depicted in Fig. \ref{fig:block_diagram}. Let $B$ denote the total number of information bits to be transmitted in a frame. A bit splitter is firstly used to split the bits into $MN/N_b$ groups, each containing $B_b=\frac{N_bB}{MN}$ bits. Every group of $B_b= b_1+b_2$ bits are mapped to the symbols in a block by an HMIM mapper, where $b_1$ bits are mapped to QAM symbols and $b_2$ bits are mapped to a mode index. To do this, we introduce the hierarchical QAM constellation (HQC) with the following definition.

\begin{definition}[Hierarchical QAM constellation]\label{definition:hqc}
    Let $\mathbf{\Lambda}\in\mathbb{C}^{Q_1\times Q_2}$ denote the set of HQC symbols, where we define the QAM order $Q_1$ and the mode order $Q_2$. Construct the set of regular $Q_1$-QAM symbols $\mathbf{\Lambda}^{\mathrm{QAM}}\in\mathbb{C}^{Q_1}$ and the set of regular $Q_2$-QAM symbols $\mathbf{\Lambda}^{\mathrm{IM}}\in\mathbb{C}^{Q_2}$. The HQC symbol corresponding to the $q_2$-th mode of the $q_1$-th QAM symbol is defined as $\Lambda[q_1,q_2]=\Lambda^{\mathrm{QAM}}[q_1]+\Lambda^{\mathrm{IM}}[q_2]$. The power ratio between $\mathbf{\Lambda}^{\mathrm{QAM}}$ and $\mathbf{\Lambda}^{\mathrm{IM}}$ is constrained by a scaling factor $\rho=d_1/d_2$, where $d_1=\min_{q_1\neq q'_1}\norm{\Lambda[q_1,q_2]-\Lambda[q'_1,q'_2]}$ and $d_2=\min_{q_2\neq q'_2}\norm{\Lambda^{\mathrm{IM}}[q_2]-\Lambda^{\mathrm{IM}}[q'_2]}$.
\end{definition}
\vspace{-5mm}
\begin{figure}[ht]
    \centering
    \includegraphics[width=8.4cm,trim={10 20 20 25},clip]{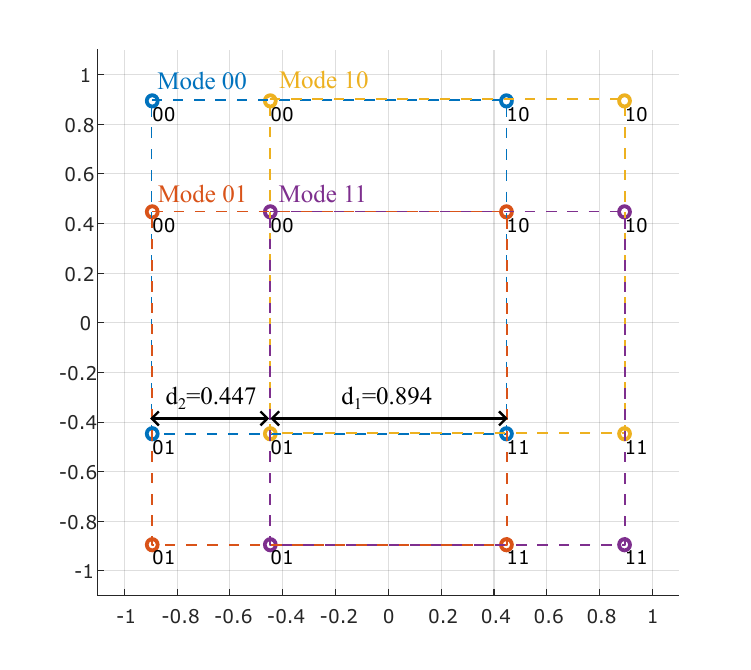}
    \vspace{-2mm}
    \caption{Normalized HQC diagram of $\mathbf{\Lambda}$ with $Q_1=4,Q_2=4,\rho=2$.}
    \label{fig:constellation}
\end{figure}

In essence, $\mathbf{\Lambda}$ is constructed from a base constellation $\mathbf{\Lambda}^{\mathrm{QAM}}$, with each QAM symbol encompassing $Q_2$ distinct modes from $\mathbf{\Lambda}^{\mathrm{IM}}$. Therefore, every $\log_2{Q_1}$ bits are mapped to a symbol in $\mathbf{\Lambda}^{\mathrm{QAM}}$ and every $\log_2{Q_2}$ bits are mapped to a mode in $\mathbf{\Lambda}^{\mathrm{IM}}$. An example of a normalized HQC with $Q_1=Q_2=4$ and $\rho = 2$ is presented in Fig. \ref{fig:constellation} with the associated information bits labeled. 

In ODDM-HMIM, we apply the same mode across $N_b$ symbols within one block to achieve a repetition gain. As shown by the example in Fig. \ref{fig:frame}, symbols in each block share one of the 4 modes labeled by the overlay integer. Let $\mathbf{\Lambda}_{q_2}$ denote the $q_2$-th column of $\mathbf{\Lambda}$. The HMIM mapper first determines the mode index $q_2$ of this block by $b_2$. Then QAM modulation is performed for individual symbols using $\mathbf{\Lambda}_{q_2}$. Therefore, we have $b_1 = N_b\log_2Q_1$ and $b_2=\log_2Q_2$. The blocks are then aggregated into a frame and transmitted by ODDM modulation. Considering the DDMC signal parameters, the SE of the ODDM-HMIM system is $\mathrm{SE} = \log_2Q_1 + (\log_2Q_2)/N_b$.

To perform ODDM modulation, the DD domain symbol frame $\mathbf{X}$ is converted to the delay-time (DT) domain by $N$-point IDFT: $\left(\mathbf{X}^{\mathrm{DT}}\right)^\tran = \mathbf{F}_N^\herm \mathbf{X}^\tran$. Then, the ODDM time-domain digital sequence $\mathbf{s}\in\mathbb{C}^{MN\times1}$ is obtained by vectorization, written as $\mathbf{s} = \mathrm{vec}(\mathbf{X}^{\mathrm{DT}})$. After that, in contrast to rectangular pulse-based OTFS, $\mathbf{s}$ is filtered by a truncated square-root Nyquist pulse $a(t)$ in ODDM to generate the continuous-time waveform:
\begin{align}
    s(t) &= \sum_{q=0}^{MN-1}s[q]a\left(t-q\frac{T}{M}\right).
    \label{eq:st}
\end{align}
Through step (\ref{eq:st}), the ODDM waveform realizes local bi-orthogonality with respect to the DD grid resolution $\mathcal{T}_\mathrm{DD}$ and $\mathcal{F}_\mathrm{DD}$ within the region $\abs{m}\leq M-1$ and $\abs{n}\leq N-1$ \cite{Lin2022OrthogonalModulation}.

\subsection{ODDM Input-Output Relation}

Consider a doubly selective channel with $P$ paths, where the delay and Doppler shifts of the $p$-th path are $\tau_p=l_p\frac{T}{M}$ and $\nu_p=k_p\frac{1}{NT}$, respectively. Also, define the sets of normalized delay and Doppler shifts by $\mathcal{L}=\{l_1,\dots,l_P\}$ and $\mathcal{K}=\{k_1,\dots,k_P\}$. Here we assume on-grid delay and Doppler shifts, i.e., $l_p,k_p\in\mathbb{Z}$.
\footnote{
    Although we assume on-grid delay and Doppler shifts for simplicity, the ODDM framework can be readily extended to off-grid channels by incorporating a sampled equivalent channel with more DD-domain taps \cite{Tong2024ODDM_PhyChan}.
}
The channel can be expressed as \cite{Bello1963CharacterizationChannels}
\begin{equation}\label{eq:ddChan_int}
    h[l,k] =
    \begin{cases}
        h_p, & l=l_p, k = k_p \\
        0, &\text{otherwise}
    \end{cases},
\end{equation}
for $l_p\in\mathcal{L},k_p\in\mathcal{K}$. The maximum delay shift is given by $l_{\max}=\max\{\mathcal{L}\}$.

When a frame-wise cyclic prefix (CP) with a length of $l_{\max}$ is deployed at the transmitter and removed at the receiver, the received signal at each time instance is the superposition of components from $\abs{\mathcal{L}}$ resolvable delay taps. Let $\mathcal{K}_l$ denote the set of Doppler shifts with nonzero response at delay shift $l$. After matched filtering and sampling at $t=q\frac{T}{M}$ for $q=0,\dots,MN-1$, the time-domain input-output relation for ODDM is derived as \cite{Tong2024ODDM_PhyChan}
\begin{equation}\label{eq:io_time_mat}
    \mathbf{r}=\mathbf{G}\mathbf{s}+\mathbf{z},
\end{equation}
where we have the received sample vector $\mathbf{r}\in\mathbb{C}^{MN\times1}$, the sampled AWGN vector $\mathbf{z}\sim\mathcal{CN}(\mathbf{0},\sigma_z^2\mathbf{I})$, and the time-domain channel matrix $\mathbf{G}\in\mathbb{C}^{MN\times MN}$ whose element at the $q$-th time index and $l$-th delay index is given by
\begin{equation}\nonumber
    G\bigl[q,[q-l]_{MN}\bigr] = \sum_{k\in{\mathcal{K}_l}} h[l,k] e^{j2\pi\frac{k(q-l)}{MN}}.
\end{equation}
Here, we define the transmit SNR as $\gamma \triangleq E_s/\sigma_z^2$, where $E_s=\mathbb{E}_q[s[q]]$ is the average signal power.

For each transmitted symbol $s[q]$, we also define a sub-input-output relation to facilitate low-complexity equalization in SIC-MMSE. To contain all the channel impaired components of $s[q]$, we construct the $q$-th sub received symbol vector $\mathbf{r}_q\in\mathbb{C}^{(l_{\max}+1)\times1}$, whose $l$-th element is $r\bigl[[q+l]_{MN}\bigr]$. Meanwhile, $\mathbf{r}_q$ is superimposed by components from interfering symbols $s\bigl[[q+\Delta l]_{MN}\bigr],\Delta l\in\dot{\mathcal{L}}$, where $\dot{\mathcal{L}}=\{-l_{\max},\dots,0,\dots,l_{\max}\}$. Following this, we also construct the $q$-th sub transmitted symbol vector $\mathbf{s}_q\in\mathbb{C}^{(2l_{\max}+1)\times1}$. Then, we define the truncated spreading vector $\mathbf{g}_{q,\Delta l}\in\mathbb{C}^{(l_{\max}+1)\times1}$ for $s\bigl[[q+\Delta l]_{MN}\bigr]$, where its $l$-th element is
\begin{equation}\label{eq:g_q_deltal}
    g_{q,\Delta l}[l] = 
    \begin{cases}
        g\bigl[l-\Delta l,[q + l]_{MN}\bigr], & l-\Delta l\in\mathcal{L},
        \\
        0, & \text{otherwise}.
    \end{cases}
\end{equation}
Therefore, we write the sub-input-output relation for $s[q]$ as
\begin{equation}\label{eq:iosub_time_mat}
    \underbrace{
    \begin{bmatrix}
        r[q]\\
        r\bigl[[q+1]_{MN}\bigr]\\
        \vdots\\
        r\bigl[[q+l_{\max}]_{MN}\bigr]
    \end{bmatrix}
    }_{\mathbf{r}_q}
    =\mathbf{G}_{q}
    \underbrace{
    \begin{bmatrix}
        s\bigl[[q-l_{\max}]_{MN}\bigr]\\
        \vdots\\
        s[q]\\
        \vdots\\
        s\bigl[[q+l_{\max}]_{MN}\bigr]
    \end{bmatrix}
    }_{\mathbf{s}_q}
    +\mathbf{z}_q,
\end{equation}
where $\mathbf{G}_{q} = [\mathbf{g}_{q,-l_{\max}},\dots,\mathbf{g}_{q,0},\dots,\mathbf{g}_{q,l_{\max}}]$ is the subchannel matrix and $\mathbf{z}_q=\left[z[q],\dots,z\bigl[[q+l_{\max}]_{MN}\bigr]\right]^\tran$ is the corresponding AWGN vector.

Additionally, by employing the inverse operation of vectorization on $\mathbf{r}$ followed by $N$-point DFT, we can convert the received signal back to the DD domain as $\mathbf{Y}\in\mathbb{C}^{M\times N}$. Regarding the discrete channel in (\ref{eq:ddChan_int}), the DD domain input-output relation of ODDM is \cite{Lin2022OrthogonalModulation,Tong2024ODDM_PhyChan}
\begin{align}
    Y[m,n] & = \sum_{p=1}^P h_p e^{j2\pi\frac{(m-l_p)k_p}{MN}} \alpha_p(m,n)
    \nonumber\\
    &\qquad \times X\bigl[[m-l_p]_M,[n-k_p]_N\bigr] + w[m,n],
    \label{eq:iodd}
\end{align}
where 
\begin{equation*}
    \alpha_p(m,n) = 
    \begin{cases}
        1, & m\geq l_p \\
        e^{-j\frac{2\pi[n-k_p]_N}{N}}, & m<l_p
    \end{cases} 
\end{equation*}
is the phase rotation for the CP symbols, and $w[m,n]$ is the DD domain noise sample. After DFT, $w[m,n]$ adheres to the same distribution as $z[q]$, i.e., $w[m,n]\sim\mathcal{CN}(0,\sigma_z^2)$.

\section{Detection Algorithms}

\subsection{ML Detector and Performance Analysis}
Considering an optimal ML detector, we analyze the BER performance of the ODDM-HMIM system, which is useful in tuning the scaling factor $\rho$. To begin with, we rewrite (\ref{eq:iodd}) as
\begin{equation}
    \mathbf{y}=\mathbf{\Phi}(\mathbf{X})\mathbf{h}+\mathbf{w},
\end{equation}
where $\mathbf{h}=[h_1,\dots,h_P]^\tran$ is the path coefficient vector, and $\mathbf{\Phi}(\mathbf{X})\in\mathbb{C}^{MN\times P}$ is the transformed symbol matrix with its element at the $(nM+m)$-th row and the $(p-1)$-th column given by
\begin{align}
    &\Phi_{mN+n,p-1}(\mathbf{X}) 
    \nonumber\\
    &\quad= \alpha_p(m,n)e^{j2\pi\frac{(m-l_p)k_p}{MN}}X\bigl[[m-l_p]_M,[n-k_p]_N\bigr].
\end{align}
Given a channel realization $\mathbf{h}$, the conditional pairwise error probability on symbol estimation $\hat{\mathbf{X}}$ is \cite{Basar2013OrthogonalModulation,Zhao2021OTFS_DMIM}
\begin{equation}\label{eq:cpep}
    P(\hat{\mathbf{X}}|\mathbf{X},\mathbf{h}) = Q\left(\sqrt{\frac{\gamma}{2}\norm*{\left(\mathbf{\Phi}\left(\hat{\mathbf{X}}\right)-\mathbf{\Phi}(\mathbf{X})\right)\mathbf{h}}^2}\right).
\end{equation}
Let $\mathbf{\Gamma} = \left(\mathbf{\Phi}\left(\hat{\mathbf{X}}\right)-\mathbf{\Phi}(\mathbf{X})\right)^\herm\left(\mathbf{\Phi}\left(\hat{\mathbf{X}}\right)-\mathbf{\Phi}(\mathbf{X})\right)$. Because $\mathbf{\Gamma}$ is Hermitian, it has the eigendecomposition $\mathbf{\Gamma} = \mathbf{U}\mathbf{\Psi}\mathbf{U}^\herm$, where $\mathbf{U}$ is unitary and the diagonal of $\mathbf{\Psi}=\diag\{\lambda_1,\dots,\lambda_P\}$ gives the eigenvalues of $\mathbf{\Gamma}$. Following equations (12)-(18) in \cite{Qian2023OTFS_BlockwiseIM}, a good approximation of the unconditional pairwise error probability with uniform path gain is given by 
\begin{equation}\label{eq:upep}
    P\left(\hat{\mathbf{X}}|\mathbf{X}\right) \approx \frac{1}{12\left(\frac{\gamma}{4P}\right)^\mathcal{R}\prod_{\mathfrak{r}}^\mathcal{R}\lambda_\mathfrak{r}} + \frac{1}{4\left(\frac{\gamma}{3P}\right)^\mathcal{R}\prod_{\mathfrak{r}}^\mathcal{R}\lambda_\mathfrak{r}},
\end{equation}
where $\mathcal{R}\leq P$ is the rank of $\Gamma$. Therefore, we can approximate the averaged BER of ODDM-HMIM by substituting (\ref{eq:upep}) to
\begin{equation}\label{eq:pb}
    P_b=\frac{1}{B}\frac{1}{Q_1^{MN}Q_2^{MN/N_b}}\sum_{\mathbf{X}}\sum_{\hat{\mathbf{X}}}P(\hat{\mathbf{X}}|\mathbf{X})e(\hat{\mathbf{X}},\mathbf{X}),
\end{equation}
where $e(\hat{\mathbf{X}},\mathbf{X})$ outputs the number of bit errors of $(\mathbf{X}\rightarrow\hat{\mathbf{X}})$.

\begin{figure}[ht]
    \centering
    \includegraphics[width=8.5cm,trim={32 4 34 22},clip]{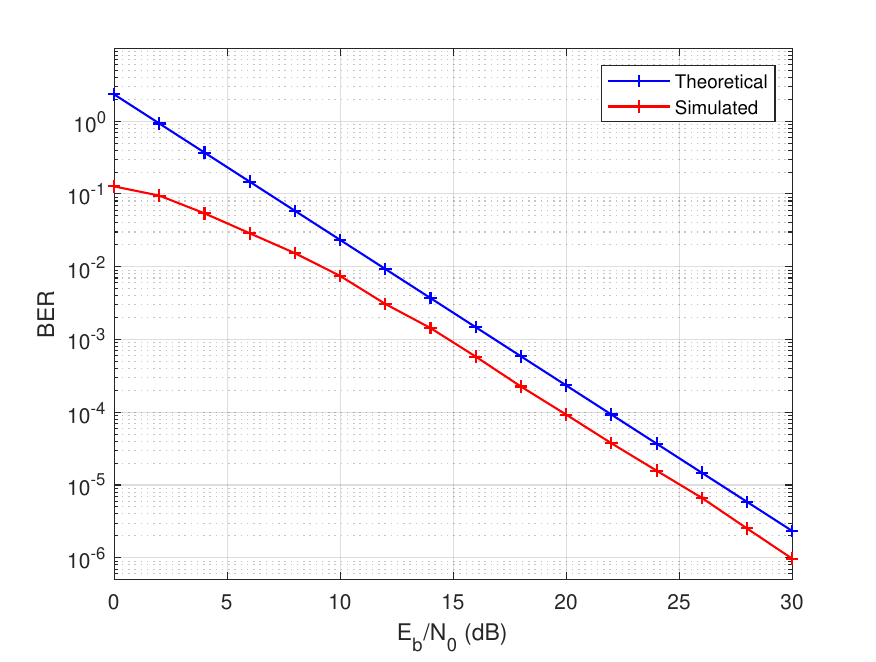}
    \vspace{-3mm}
    \caption{BER performance of ODDM-HMIM with ML detection (SE = 1.5 bps/Hz, $M=N=2$, $P=2$).}
    \label{fig:ml_analysis}
    \vspace{-2mm}
\end{figure}

The derived equation for BER is examined at ODDM frame size $M=N=2$ with HMIM parameters $Q_1=Q_2=N_b=2,\rho=1.4$. The analytical and simulated BER results are plotted in Fig. \ref{fig:ml_analysis}. It is observed that our approach can closely approximate the actual BER results, especially at high SNR values. We then perform incremental searches based on (\ref{eq:pb}) to numerically optimize the scaling factor $\rho$.

\vspace{-3mm}

\subsection{MAP Estimator under Gaussian noise}\label{sec:map}
\begin{figure}[ht]
    \centering
    \includegraphics[width=4.5cm,trim={0 0 0 0},clip]{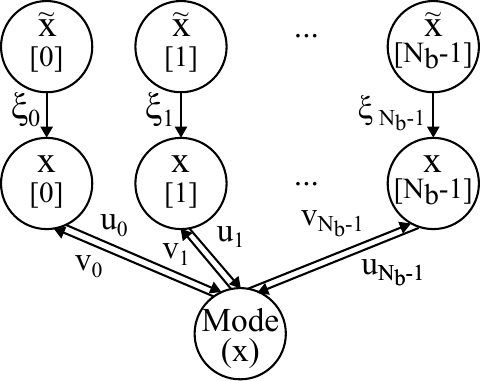}
    \vspace{-2mm}
    \caption{Factor graph of the block-wise MAP estimator for ODDM-HMIM with Gaussian noise.}
    \label{fig:factor_graph}
\end{figure}

In this section, we derive the MAP estimator for ODDM-HMIM by treating interference as \emph{Gaussian} noise. Equivalently, we consider a single transmitted block $\mathbf{x}_{m,\beta}$. In this section, we drop the block indexes and denote the block as $\mathbf{x}$. The block observation is modelled as $\tilde{\mathbf{x}} = \mathbf{x} + \mathbf{\Delta}$ with interference plus noise $\mathbf{\Delta}\sim\mathcal{CN}(\mathbf{0},\sigma_\Delta^2\mathbf{I})$. The MAP estimator for $\mathbf{x}$ is written as $\hat{\mathbf{x}} = \argmax_{\mathbf{x}} P(\mathbf{x}|\tilde{\mathbf{x}})$, where
\begin{align}
    P(\mathbf{x}|\tilde{\mathbf{x}}) &= P(\tilde{\mathbf{x}}|\mathbf{x})P(\mathbf{x})/P(\tilde{\mathbf{x}})
    \nonumber\\
    &= P(\tilde{\mathbf{x}}|\mathbf{x})P(\mathbf{x}|\mathrm{Mode}(\mathbf{x}))P(\mathrm{Mode}(\mathbf{x}))/P(\tilde{\mathbf{x}}),
    \label{eq:map_og}
\end{align}
and $\mathrm{Mode}(\mathbf{x})$ outputs the mode index of the block $\mathbf{x}$. Since the symbols in $\mathbf{x}$ are i.i.d., (\ref{eq:map_og}) can be decomposed as
\begin{align}
    P(\mathbf{x}|\tilde{\mathbf{x}}) &= P(\mathrm{Mode}(\mathbf{x}))\prod_{n_b}\bigl(1/P(\tilde{x}[n_b])\bigr) 
    \nonumber\\
    &\quad \times\prod_{n_b}\bigl(P(\tilde{x}[n_b]|x[n_b])P(x[n_b]|\mathrm{Mode}(\mathbf{x}))\bigr),
    \label{eq:map}
\end{align}
with the factor graph representation in Fig. \ref{fig:factor_graph}.

The message from $\tilde x[n_b]$ to $x[n_b]$ is the symbol likelihood with respect to the HQC $\mathbf{\Lambda}$:
\begin{align}\label{eq:xi}
    \xi_{n_b}[q_1,q_2] = \frac{\exp\left(\frac{-\abs{\tilde{x}[n_b]-\Lambda[q_1,q_2]}^2}{\sigma_\Delta^2}\right)}{\sum_{q'_1,q'_2}\exp\left(\frac{-\abs{\tilde{x}[n_b]-\Lambda[q'_1,q'_2]}^2}{\sigma_\Delta^2}\right)}.
\end{align}
The mode likelihood from $x[n_b]$ to the $\mathrm{Mode}(\mathbf{x})$ is
\begin{equation}\label{eq:u}
    u_{n_b}[q_2] = \sum_{q_1}\xi_{n_b}[q_1,q_2].
\end{equation}
Then, the extrinsic information from $\mathrm{Mode}(\mathbf{x})$ to $x[n_b]$ is
\begin{align}\label{eq:v}
    v_{n_b}[q_2] = \frac{\prod_{n'_b\neq n_b} u_{n'_b}[q_2]}{\sum_{q'_2}\prod_{n'_b\neq n_b} u_{n'_b}[q'_2]}.
\end{align}
The probability mass function (PMF) of symbol estimation can be extracted as
\begin{equation}\label{eq:map_pmf}
    P(x[n_b]=\Lambda[q_1,q_2]|\tilde{\mathbf{x}}) = \frac{\xi_{n_b}[q_1,q_2]v_{n_b}[q_2]}{\sum_{q'_1,q'_2}\xi_{n_b}\left[q'_1,q'_2\right]v_{n_b}[q'_2]}.
\end{equation}
Likewise, the PMF of the block mode is
\begin{equation}\label{eq:map_pmf_mode}
    P(\mathrm{Mode}(\mathbf{x})=q_2|\tilde{\mathbf{x}}) = \frac{\prod_{n_b} u_{n_b}[q_2]}{\sum_{q'_2}\prod_{n_b} u_{n_b}[q'_2]}.
\end{equation}

Here, equations (\ref{eq:xi}) and (\ref{eq:map_pmf}) have the dominant computational complexity. Therefore, the MAP estimation for an HMIM block has a linear complexity order of $\mathcal{O}(N_bQ_1Q_2)$. By comparison, the MAP/ML estimation for such a block in the conventional IM scheme \cite{Liang2020DopplerModulation} requires brute-force search among all the realizations and yields a complexity order of $\mathcal{O}\left(Q^{K_b}\binom{N_b}{K_b}\right)$, where $Q$ is the modulation order of regular $Q$-QAM and $K_b$ is the number of active indexes. In the presence of ISI, this low-complexity MAP estimator can be combined with an equalizer to perform signal detection, such as within the SIC-MMSE detector presented in the next section.

\subsection{SIC-MMSE Detector}\label{sec:sicmmse}
Originally proposed in \cite{Li2024SICMMSE_Turbo}, the SIC-MMSE detector operates in both the time domain and the DD domain. Modifications are made to align with the ODDM-HMIM scheme. The estimation of each symbol starts with canceling the interference components from the time-domain received signal. Denote prior symbol estimates by $\hat{s}[q] = s[q] + \Delta s[q],q=0,\dots,MN-1$ with $\Delta s[q]$ being the symbol estimation error. Without prior information, the means and variances of all the symbol estimates are simply zero and $E_s$, respectively. And define $\tilde{\mathbf{r}}_q\in\mathbb{C}^{(l_{\max}+1)\times1}$ to be the channel impaired branch vector for $s[q]$. Referring to the time-domain sub-input-output relation in (\ref{eq:iosub_time_mat}), we can perform interference cancellation and obtain $\tilde{\mathbf{r}}_q = \mathbf{g}_{q,0} s[q] + \tilde{\mathbf{z}}_q$, where
\begin{equation}\label{eq:ripn}
    \tilde{\mathbf{z}}_q = \mathbf{z}_q -\sum_{\Delta l\neq0,\Delta l\in\dot{\mathcal{L}}} \mathbf{g}_{q,\Delta l}\Delta s\bigl[[q+\Delta l]_{MN}\bigr]
\end{equation}
is the residual interference plus noise (RIPN) vector.

Then, MMSE equalization for the $q$-th time-domain symbol can be performed as $\tilde{s}[q] = \mathbf{w}_q\tilde{\mathbf{r}}_q/\mu_q$, where
\begin{equation}\label{eq:mmse_filter}
    \mathbf{w}_q = \mathbf{g}_q^\herm\left(\mathbf{G}_q\mathbf{V}_q\mathbf{G}_q^\herm+\sigma_z^2\right)^{-1},
\end{equation}
is the MMSE filter, and $\mu_q=\mathbf{w}_q\mathbf{g}_q$ is the normalization factor. We have $\mathbf{V}_q=\mathbb{E}\left[\Delta\mathbf{s}_q\Delta\mathbf{s}_q^\herm\right]$ being the covariance matrix of the symbol estimates. After performing $N$-point DFT on the time-domain equalized symbols, we use $\tilde{\mathbf{x}}_m$ to denote the obtained DD-domain equalized symbol vector at the $m$-th delay. Following equation (28) in \cite{Li2024SICMMSE_Turbo}, the post-MMSE error is assumed to be Gaussian and we can obtain its approximated variance $\var(\Delta\tilde{x}_m[n])$ for $n=0,\dots,N-1$.

Utilizing the DD-domain MMSE output $\tilde{\mathbf{x}}_m$, we exploit the constellation constraints of ODDM-HMIM by performing block-wise soft symbol estimation, which diverges from the original SIC-MMSE algorithm in \cite{Li2024SICMMSE_Turbo}. With the Gaussian error assumption and using the MAP estimator derived in section \ref{sec:map}, we obtain the PMF $P(x_{m}[n]=\Lambda[q_1,q_2]|\tilde{\mathbf{x}}_{m,\beta})$ by (\ref{eq:map_pmf}). Thus, we have the a posteriori mean of $x_m[n]$
\begin{align}
    \hat{x}_m[n] &=\sum_{q_1,q_2} P(x_m[n]=\Lambda[q_1,q_2]|\tilde{\mathbf{x}}_{m,\beta})\Lambda[q_1,q_2],
\end{align}
and the corresponding a posteriori error variance
\begin{align}
    &\var(\Delta x_m[n])=
    \nonumber\\
    &\sum_{q_1,q_2} P(x_m[n]=\Lambda[q_1,q_2]|\tilde{\mathbf{x}}_{m,\beta})\abs*{\Lambda[q_1,q_2]-\hat{x}_m[n]}^2.
\end{align}
Next, $\hat{x}_m[n]$ is converted back to the time domain by $N$-point IDFT. The corresponding time-domain covariance matrix is obtained following \cite{Li2024SICMMSE_Turbo} equation (34). In subsequent interference cancellation and symbol estimation, the a posteriori estimates and variances will be used as prior information to perform SIC-MMSE equalization.

The SIC-MMSE detector runs iteratively to harvest gain from prior symbol estimates. The converged symbol PMF and mode PMF can be extracted in the last iteration by (\ref{eq:map_pmf}) and (\ref{eq:map_pmf_mode}), respectively. As the covariance matrix $\mathbf{V}_q$ being updated continuously, the SIC-MMSE filter $\mathbf{w}_q$ is recomputed by (\ref{eq:mmse_filter}) in every iteration. The associated matrix inversion yields a complexity order of $\mathcal{O}\left((l_{\max}+1)^3\right)$, becoming a dominant source of computational complexity. Therefore, considering DFT, IDFT, and the MAP estimator given in section \ref{sec:map}, the overall complexity order of the soft SIC-MMSE algorithm is $\mathcal{O}\left(n_\mathrm{ite}MN\left(Q_1Q_2+\log_2N+(l_{\max}+1)^3\right)\right)$.

Remarkably, other low-complexity detectors, including the maximal-ratio combining (MRC) detector \cite{Thaj2020LowSystems} and the hard-decision SIC-MMSE detector \cite{Li2024SICMMSE_Turbo}, are also applicable to ODDM-HMIM by invoking the MAP estimator in section \ref{sec:map}. However, the soft-decision SIC-MMSE detector described in this paper demonstrates the best BER performance.

\section{Numerical Results}\label{sec:numerical_results}

In this section, we present the numerical results for the BER performance of the ODDM-HMIM system. A carrier frequency of 5 GHz and a subcarrier spacing of 15 kHz are used. The modulation parameters used for different SEs are summarized in Table \ref{tab:se}. For doubly selective channels, random Doppler shifts are generated for paths by the Jakes' model \cite{Jakes1974MicrowaveCommunications} with a user equipment (UE) speed of 500 km/h. More than 100 frame errors are collected for each simulation, ensuring a robust reflection of the BER performance.

\vspace{-3mm}
\begin{table}[ht]
    \centering
    \caption{Modulation Parameters for Different SEs}
    \vspace{-2mm}
    \label{tab:se}
    \begin{tabular}{cccl}
        \toprule
        {SE (bps/Hz)} & {ODDM} & {ODDM-IM} & {ODDM-HMIM}\\
        \midrule
        \multirow{2}{*}{2} & \multirow{2}{*}{4-QAM} & {4-QAM} &{$Q_1=Q_2=2$}\\
        {} & {} & {$N_b=1,K_b=3$} & {$N_b=1,\rho=1$}\\
        \midrule
        \multirow{2}{*}{2.5} & \multirow{2}{*}{-} & {16-QAM} &{$Q_1=Q_2=4$}\\
        {} & {} & {$N_b=1,K_b=2$} & {$N_b=4,\rho=1.1$}\\
        \midrule
        \multirow{2}{*}{3} & \multirow{2}{*}{8-QAM} & \multirow{2}{*}{-} &{$Q_1=Q_2=4$}\\
        {} & {} & {} & {$N_b=2,\rho=1.1$}\\
        \bottomrule
    \end{tabular}
\end{table}

\begin{figure}[ht]
    \centering
    \includegraphics[width=8.5cm,trim={32 4 34 18},clip]{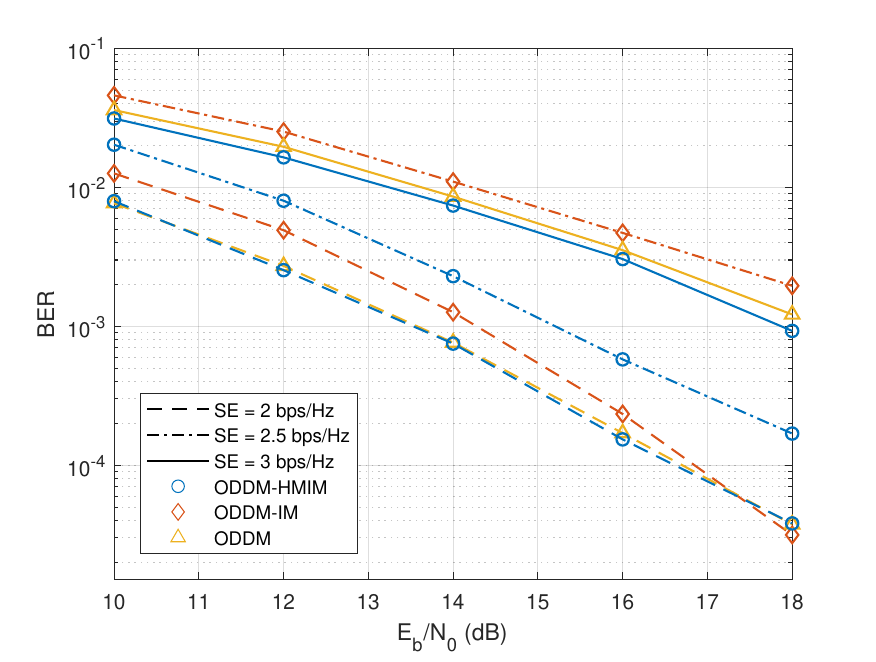}
    \vspace{-2mm}
    \caption{Performance comparison of ODDM-HMIM, ODDM-IM, and, ODDM ($M=N=32$).}
    \label{fig:vs_oddm_im}
    \vspace{-4mm}
\end{figure}

The BER performance of the proposed ODDM-HMIM system is firstly compared with ODDM and ODDM-IM systems, as presented in Fig. \ref{fig:vs_oddm_im}. Here, the ODDM-IM system is established by incorporating the IM scheme from \cite{Liang2020DopplerModulation} into an ODDM system. All the systems use linear MMSE equalization followed by block-wise ML detection, which is common in OTFS-IM literatures \cite{Liang2020DopplerModulation,Zhao2021OTFS_DMIM,Li2022OTFS_IM_EP}, so the simulation is performed with a small frame size $M=N=32$. The channel has a uniform power delay profile with 5 taps. At SE = 2 bps/Hz, ODDM-HMIM outperforms ODDM-IM at most of the $E_b/N_0$ region. Notably, because HMIM with $Q_1=Q_2=2,N_b=1$ reduces to regular 4-QAM, it can be seen that the BER of ODDM and ODDM-HMIM are identical with a SE of 2 bps/Hz. When the SE increases to 2.5 bps/Hz, the performance gap between ODDM-HMIM and ODDM-IM becomes larger. This is because ODDM-HMIM realizes a middle ground between 4-QAM and 16-QAM, while ODDM-IM needs to use 16-QAM on individual symbols to achieve such a SE. ODDM-HMIM also has a performance gain over ODDM at SE = 3 bps/Hz, where we use a non-square 8-QAM from MATLAB \cite{qammod} for ODDM. On the other hand, ODDM-HMIM fully utilizes the modulation space, achieving consistent BER performance at various SEs.

\begin{figure}[ht]
    \centering
    \includegraphics[width=8.5cm,trim={32 4 34 18},clip]{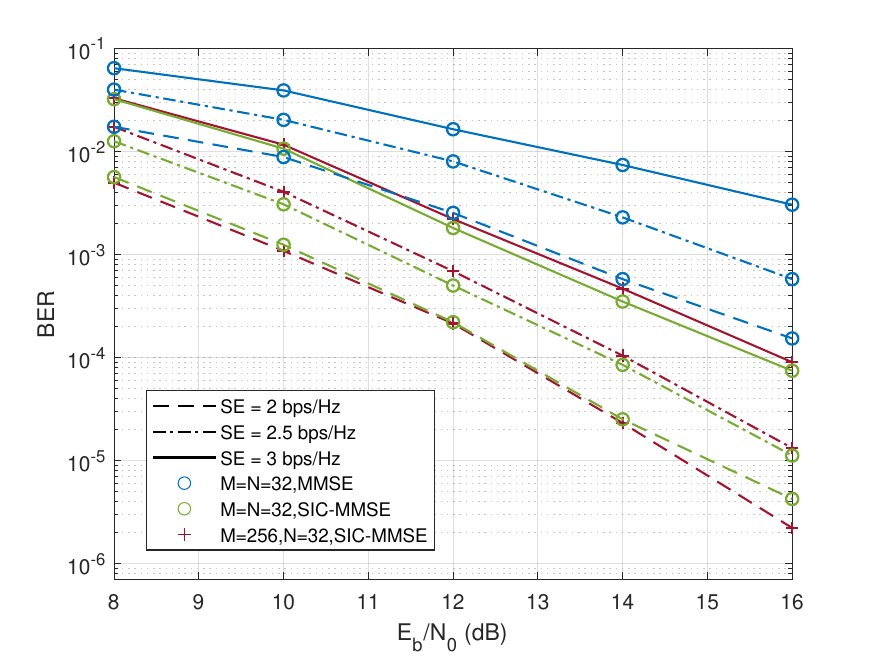}
    \vspace{-2mm}
    \caption{BER performance of ODDM-HMIM with MMSE and SIC-MMSE.}
    \label{fig:sicmmse}
\end{figure}

We then apply the ODDM system model and evaluate the performance of the SIC-MMSE detector, which is presented in Fig. \ref{fig:sicmmse}. At frame size $M=N=32$, it can be observed that the SIC-MMSE detector has a significant gain in error performance over the linear MMSE detector. This is because the SIC-MMSE detector can iteratively harvest the decision gain through the MAP estimator. Notably, this improvement is achieved without a significant increase in computational complexity because the sub-input-output relation in (\ref{eq:iosub_time_mat}) is used. We also performed simulations with a large frame size $M=256,N=32$, where the power delay profile of the EVA model is adopted. In this scenario, linear MMSE has prohibitive complexity. However, the SIC-MMSE detector still operates efficiently and approaches the BER performance with $M=N=32$ across different SEs.

\vspace{-2mm}
\vspace{-2mm}

\section{Conclusion}\label{sec:conclusion}

In this paper, we introduce the ODDM-HMIM scheme, offering a significant reduction in computational complexity while achieving better SE tradeoffs over conventional IM schemes. With the simplified IM mapping method, HMIM realizes linear computational complexity in MAP detection under Gaussian noise. It is also observed that, compared to ODDM and ODDM-IM, ODDM-HMIM shows comparable or superior BER performance, especially at high SE. Furthermore, we introduce the SIC-MMSE detector for ODDM-HMIM, which significantly improves BER performance and reduces detection complexity compared to the conventional linear MMSE detector. These advancements of ODDM-HMIM contribute to the efficiency and accuracy of signal detection in DD domain communications over doubly selective channels.


%






\ifCLASSOPTIONcaptionsoff
  \newpage
\fi



\bibliographystyle{IEEEtran}
\bibliography{references}

\begin{thebibliography}{10}
\providecommand{\url}[1]{#1}
\csname url@samestyle\endcsname
\providecommand{\newblock}{\relax}
\providecommand{\bibinfo}[2]{#2}
\providecommand{\BIBentrySTDinterwordspacing}{\spaceskip=0pt\relax}
\providecommand{\BIBentryALTinterwordstretchfactor}{4}
\providecommand{\BIBentryALTinterwordspacing}{\spaceskip=\fontdimen2\font plus
\BIBentryALTinterwordstretchfactor\fontdimen3\font minus \fontdimen4\font\relax}
\providecommand{\BIBforeignlanguage}[2]{{%
\expandafter\ifx\csname l@#1\endcsname\relax
\typeout{** WARNING: IEEEtran.bst: No hyphenation pattern has been}%
\typeout{** loaded for the language `#1'. Using the pattern for}%
\typeout{** the default language instead.}%
\else
\language=\csname l@#1\endcsname
\fi
#2}}
\providecommand{\BIBdecl}{\relax}
\BIBdecl

\bibitem{Hadani2017OrthogonalModulation}
R.~Hadani, S.~Rakib, M.~Tsatsanis, A.~Monk, A.~J. Goldsmith, A.~F. Molisch, and R.~Calderbank, ``Orthogonal time frequency space modulation,'' in \emph{Proc. IEEE WCNC}, 2017, pp. 1--6.

\bibitem{Raviteja2018InterferenceModulation}
P.~Raviteja, K.~T. Phan, Y.~Hong, and E.~Viterbo, ``Interference cancellation and iterative detection for orthogonal time frequency space modulation,'' \emph{IEEE Trans. Wireless Commun.}, vol.~17, no.~10, pp. 6501--6515, 2018.

\bibitem{Lin2022OrthogonalModulation}
H.~Lin and J.~Yuan, ``Orthogonal delay-{Doppler} division multiplexing modulation,'' \emph{IEEE Trans. Wireless Commun.}, vol.~21, no.~12, pp. 11\,024--11\,037, 2022.

\bibitem{Thaj2020LowSystems}
T.~Thaj and E.~Viterbo, ``Low complexity iterative rake decision feedback equalizer for zero-padded {OTFS} systems,'' \emph{IEEE Trans. Veh. Technol.}, vol.~69, no.~12, pp. 15\,606--15\,622, 2020.

\bibitem{Tong2024ODDM_PhyChan}
J.~Tong, J.~Yuan, H.~Lin, and J.~Xi, ``Orthogonal delay-{Doppler} division multiplexing ({ODDM}) over general physical channels,'' \emph{IEEE Trans. Commun.}, pp. 1--1, 2024.

\bibitem{Lin2022OnPulse}
H.~Lin and J.~Yuan, ``On delay-{Doppler} plane orthogonal pulse,'' in \emph{Proc. IEEE GLOBECOM}, 2022, pp. 5589--5594.

\bibitem{Huang2024ODDM_Performance}
K.~Huang, M.~Qiu, J.~Tong, J.~Yuan, and H.~Lin, ``Performance of orthogonal delay-{Doppler} division multiplexing modulation with imperfect channel estimation,'' \emph{IEEE Trans. Commun.}, pp. 1--1, 2024.

\bibitem{Basar2013OrthogonalModulation}
E.~Basar, U.~Aygolu, E.~Panayirci, and H.~V. Poor, ``Orthogonal frequency division multiplexing with index modulation,'' \emph{IEEE Trans. Signal Process.}, vol.~61, no.~22, pp. 5536--5549, 2013.

\bibitem{Liang2020DopplerModulation}
Y.~Liang, L.~Li, P.~Fan, and Y.~Guan, ``Doppler resilient orthogonal time-frequency space ({OTFS}) systems based on index modulation,'' in \emph{Proc. IEEE VTC}, 2020, pp. 1--5.

\bibitem{Zhao2021OTFS_DMIM}
H.~Zhao, D.~He, Z.~Kang, and H.~Wang, ``Orthogonal time frequency space ({OTFS}) with dual-mode index modulation,'' \emph{IEEE Wireless Commun. Lett.}, vol.~10, no.~5, pp. 991--995, 2021.

\bibitem{Qian2023OTFS_BlockwiseIM}
M.~Qian, F.~Ji, Y.~Ge, M.~Wen, X.~Cheng, and H.~V. Poor, ``Block-wise index modulation and receiver design for high-mobility {OTFS} communications,'' \emph{IEEE Trans. Commun.}, vol.~71, no.~10, pp. 5726--5739, 2023.

\bibitem{Zhang2017OFDM_DMIM}
X.~Zhang, H.~Bie, Q.~Ye, C.~Lei, and X.~Tang, ``Dual-mode index modulation aided {OFDM} with constellation power allocation and low-complexity detector design,'' \emph{IEEE Access}, vol.~5, pp. 23\,871--23\,880, 2017.

\bibitem{Li2022OTFS_IM_EP}
L.~Li, L.~Xiao, X.~Zhang, G.~Liu, S.~Li, and T.~Jiang, ``An efficient symbol-by-symbol aided expectation propagation detector for {OTFS} with index modulation,'' \emph{IEEE Wireless Commun. Lett.}, vol.~11, no.~10, pp. 2046--2050, 2022.

\bibitem{Bello1963CharacterizationChannels}
P.~A. Bello, ``Characterization of randomly time-variant linear channels,'' \emph{IEEE Trans. Commun. Syst.}, vol.~11, no.~4, pp. 360--393, 1963.

\bibitem{Li2024SICMMSE_Turbo}
Q.~Li, J.~Yuan, M.~Qiu, S.~Li, and Y.~Xie, ``Low complexity turbo {SIC-MMSE} detection for orthogonal time frequency space modulation,'' \emph{IEEE Trans. Commun.}, vol.~72, no.~6, pp. 3169--3183, 2024.

\bibitem{Jakes1974MicrowaveCommunications}
W.~C. Jakes, \emph{Microwave Mobile Communications}.\hskip 1em plus 0.5em minus 0.4em\relax Wiley-IEEE Press, 1974.

\bibitem{qammod}
\BIBentryALTinterwordspacing
MathWorks, ``qammod,'' 2024. [Online]. Available: \url{https://au.mathworks.com/help/comm/ref/qammod.html}
\BIBentrySTDinterwordspacing

\end{thebibliography}

\end{document}